\def\hst{{\sl HST}}

\def\spitzer{{\sl Spitzer}}
\def\irac{{\sl IRAC}}

\def\gc {{GC}}

\documentclass[11pt,twoside]{article}
\usepackage{asp2006}
\usepackage{psfig}
\usepackage{epsfig}
\usepackage{epsf}
\usepackage{graphics}
\usepackage{lscape}
\def\edcomment#1{\iffalse\marginpar{\raggedright\sl#1\/}\else\relax\fi}
\reversemarginpar \textwidth 410pt %\oddsidemargin -0.1in

\begin{document}
\title{HST Pa$\alpha$ Survey of the Galactic Center -- Searching the missing young stellar populations within the Galactic Center}
\author{H. Dong$^{1}$, Q. D. Wang$^1$, A. Cotera$^2$, S.
Stolovy$^3$, M. R. Morris$^4$, J. Mauerhan$^3$, E. A. Mills$^4$,
G. Schneider$^5$, C. Lang$^6$}

\affil{$^1$ Department of Astronomy, University of Massachusetts,
Amherst, MA 01003} \affil{$^2$ SETI Institute} \affil{$^3$ Spitzer
Science Center, California Institute of Technology} \affil{$^4$
Department of Physics and Astronomy, University of California, Los
Angeles}\affil{$^5$ Steward Observatory, University of Arizona}
\affil{$^6$ Department of Physics and Astronomy, University of
Iowa}
\affil{E-mail: hdong@astro.umass.edu, wqd@astro.umass.edu}

\begin{abstract}
We present preliminary results of our \hst\ Pa$\alpha$ survey of
the Galactic Center (\gc ), which maps the central
0.65$\times$0.25 degrees around Sgr A*. This survey provides us
with a more complete inventory of massive stars within the \gc ,
compared to previous observations. We find 157 Pa$\alpha$ emitting sources, which are evolved massive
stars. Half of them are located outside of three young massive
star clusters near Sgr A*. The loosely spatial distribution of
these field sources suggests that they are within less massive
star clusters/groups, compared to the three massive ones. Our
Pa$\alpha$ mosaic not only resolves previously well-known
large-scale filaments into fine structures, but also reveals many
new extended objects, such as bow shocks and H II regions. In
particular, we find two regions with large-scale Pa$\alpha$
diffuse emission and tens of Pa$\alpha$ emitting sources in the
negative Galactic longitude suggesting recent star formation
activities, which were not known previously. Furthermore, in our
survey, we detect $\sim$0.6 million stars, most of which are red
giants or AGB stars. Comparisons of the magnitude distribution in
1.90 $\mu$m and those from the stellar evolutionary tracks with
different star formation histories suggest an episode of star
formation process about 350 Myr ago in the \gc .
\end{abstract}

\vspace{-0.5cm}
\section{Motivation}\label{s:motivation}
As the closest galactic nucleus, the GC (8kpc, [1]) is a unique
lab to study resolved stellar populations around a supermassive
black hole (SMBH). This topic was triggered by the discovery of
three young massive compact star clusters within 30 pc around Sgr
A* in the last two decades ($<$ 6 Myr old, $\sim10^4~M_{\odot}$,
[2, 3, 4] and references therein). These clusters contribute
strong UV radiation that illuminates nearby molecular clouds. The
existence of these clusters provides us with a good opportunity to
understand the star formation process within a galactic nucleus,
which is very important to the study of galaxy formation and
evolution.

However, the formation mechanism of these clusters ([5,6]) still
remains greatly uncertain, under the hostile GC environment,
characterized by high temperature gas as well as strong magnetic
field and tidal force. Existing studies are focused on the Arches
and Center clusters ([7,8]). This small sample of massive stellar
clusters prevents us from making a firm conclusion about their
formation mode. We do not even know whether they were formed in a
single burst or were just part of a continuous star formation in
the GC. We are also not sure about whether or not these extreme
massive star clusters represent the only way in which stars form
in the \gc. A survey of a fair sample of massive stars will thus
be very helpful.

We also need to determine the long-term star formation history in
the GC. Figer et al 2004 ([9]) first studied the K-band luminosity
function toward the \gc\ obtained by the HST/NICMOS NIC2. They
found that a continuous star formation is consistent with their
data. Their conclusion, however, is based on stars detected in
several regions with small fields of view. Existing large-scale
near-IR surveys, such as 2MASS ([11]), are dominated by foreground
stars, plus a limited number of bright red giants and AGB stars in
the \gc. Therefore, it is highly desirable to have a large-scale,
high-resolution near-IR survey, which will facilitate a study of
differential properties of the luminosity function across the GC.

We have carried out a HST/NICMOS Pa$\alpha$ survey of the Galactic
Center ([10]). This survey used the NIC3, which has a relatively
large field of view (51.2"$\times$51.2") and an angular resolution
($\sim$0.2"). This combination, together with the stable PSF,
allows us to effectively separate the point sources contribution
from the extended diffuse Pa$\alpha$ emission. Arising chiefly
from photo-ionized warm gas, Pa$\alpha$ line emission (at
1.87$\mu$m) is brighter than Br$\gamma$ (2.16$\mu$m, accessible
from the ground) by a factor of 3-4, even considering the high
extinction toward the GC. In particular, point-like Pa$\alpha$
emission should be predominantly produced by evolved massive stars
(e.g., O If, LBV and WR) with age younger than $\sim$10 Myr.

\section{Observation and Data Reduction}\label{s:observation}
Our survey covered the central 416 square arcminutes of
the GC in 144 \hst\ orbits. Two narrow-band filters, F187N (1.87
$\mu$m, on line) and F190N (1.90 $\mu$m, off line) were used with the same
orbit parameters. Each orbit included four pointing positions.
Each was observed with a four-point dithering pattern.
The exposure time is 192 s for each position and filter.

We will present in Dong et al 2010 ([12]) a detailed description
of the data calibration and analysis procedures. Here, we list
several key steps. We remove the relative background offsets among
the 576 positions simultaneously determined from the detected
intensity differences in the overlap regions. The absolute
background is determined and removed based on the intensities
observed toward several foreground dark molecular clouds. The
relative spatial offsets among the orbits are corrected for in the
same fashion, while the absolute astrometry is determined from a
comparison with the accurate positions of the SiO masers in Reid
et al 2007.

We use the IDL routine 'Starfinder' developed by Emiliano Diolaiti
([13]) to detect stars. The 50\% completeness limit of the
detection is typically $m_{1.90 \mu m}$=17 and is reduced to
$m_{1.90 \mu m} \sim 15$ magnitude near Sgr A*, where faint source
confusion becomes important. We are also able to quantify the
photometric errors, accounting for the Poisson uncertainty, the
background fluctuation,  and the NIC3's inter-pixel problem. In
total, we detect $\sim$0.6 million sources at 5$\sigma$ confidence
in both filters.

We then produce an extinction map calculated adaptively from local
flux ratios of sources in the two filters. We use this extinction
map to correct for the spatial variable extinction of the
Pa$\alpha$ emission. We further excise individual Pa$\alpha$
sources to construct a diffuse Pa$\alpha$ emission map, which is
present in Fig.~\ref{f:total_color}, together with the final F187N
and F190N mosaics.

\section{Result}\label{s:result}
\subsection{Magnitude Distribution}\label{s:magnitude}
In the left panel of Fig.~\ref{f:point}, we present a magnitude
distribution of the 0.6 million sources in F190N. This
distribution shows two peaks at $\sim$15.5 and 17 magnitudes.
While the later peak could be explained by the variable
incompleteness of the source detection, the former one should be
real, which cannot be due to extinction variation across the
field, for example. To understand the nature of the 15.5 magnitude
peak, we compare the F190N magnitude contours with the Padova
stellar evolutionary tracks in the right panel of
Fig.~\ref{f:point}. The F190N contours are calculated from the
ATLAS 9 atmosphere model and corrected for the distance and
extinction ($A_{V}=31.1$). As shown by the figure, most of the
detected sources should be red giants and AGB stars. Only Main
Sequence stars more massive than $\sim 9 M_{\odot}$ should be
detected individually in our survey. The peak around 15.5
magnitude is apparently due to red clump stars with initial mass
around 1-3 $M_{\odot}$.

\begin{figure*}[!thb]
  \centerline{
    \epsfig{figure=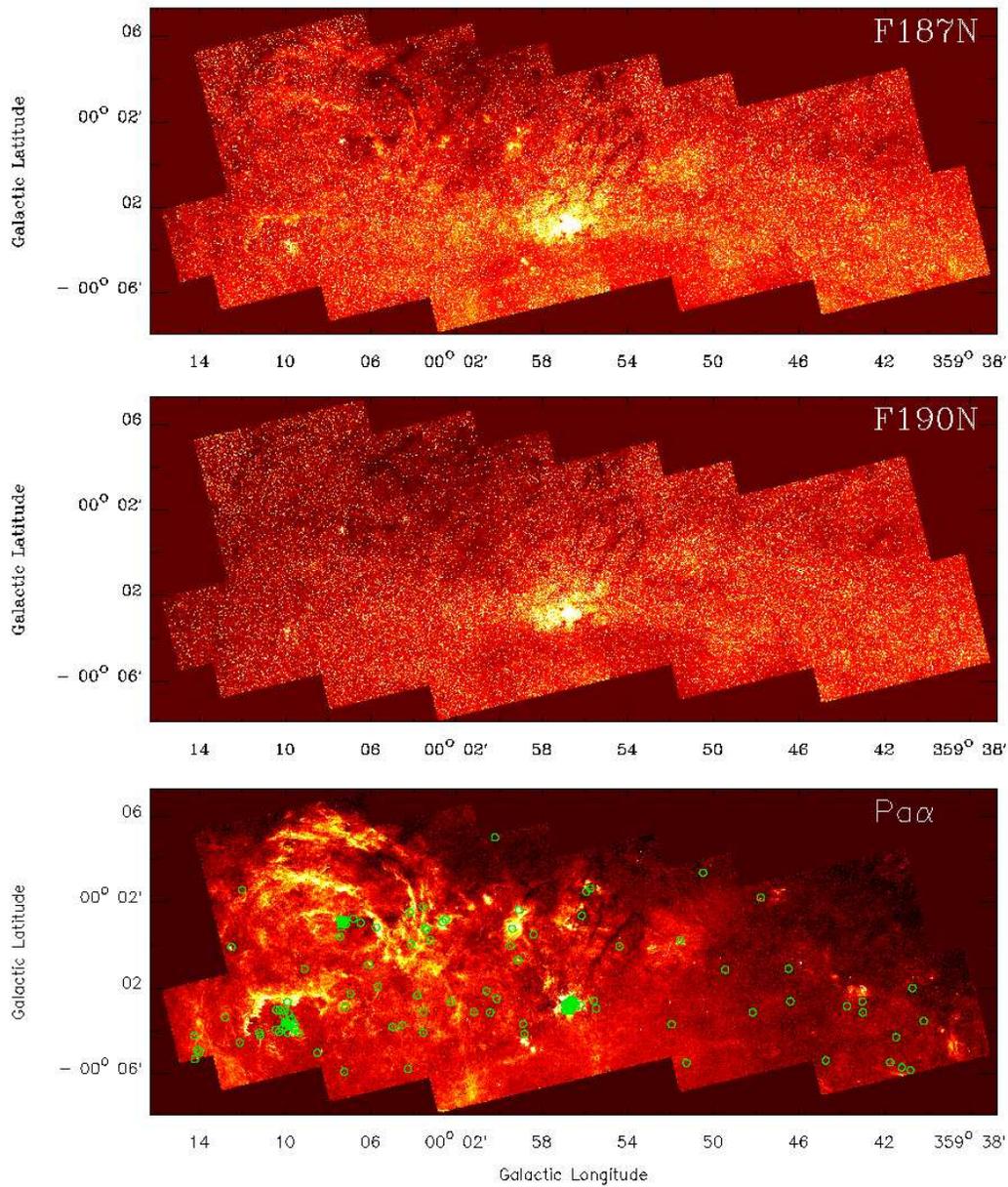,width=1.0\textwidth,angle=0}
    }
 \caption{The mosaics from our survey:
 F187N (top), F190N (middle) and diffuse Pa$\alpha$ emission (bottom).
 The green circles in
 the Pa$\alpha$ mosaic mark the positions of the detected Pa$\alpha$
 emitting sources (see Section 6).
 }
 \label{f:total_color}
\end{figure*}

\begin{figure*}[!thb]
  \centerline{
    \epsfig{figure=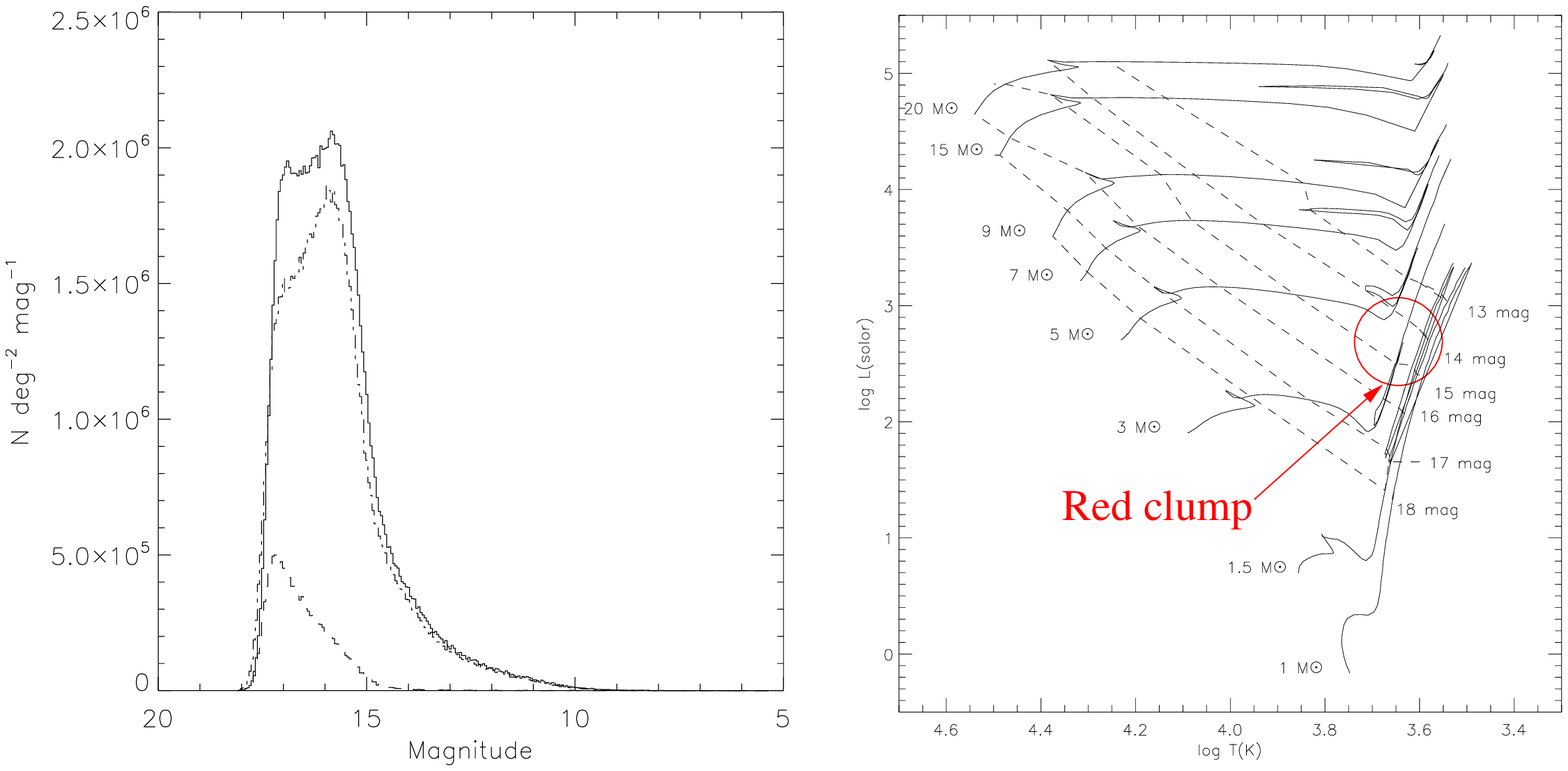,width=1.0\textwidth,angle=0}
    }
 \caption{Left: The magnitude distribution of f190N
sources. The solid histogram represents the magnitude
distributions of all sources, while the dashed and dash-dotted
ones are for the foreground (defined to have the color H-K$<$1 )
and the remaining (GC) sources, separately. Right: The Padova
stellar evolutionary tracks with the
 F190 magnitude contours are overlaid.
}
\label{f:point}
\end{figure*}

\begin{figure*}[!thb]
  \centerline{
    \epsfig{figure=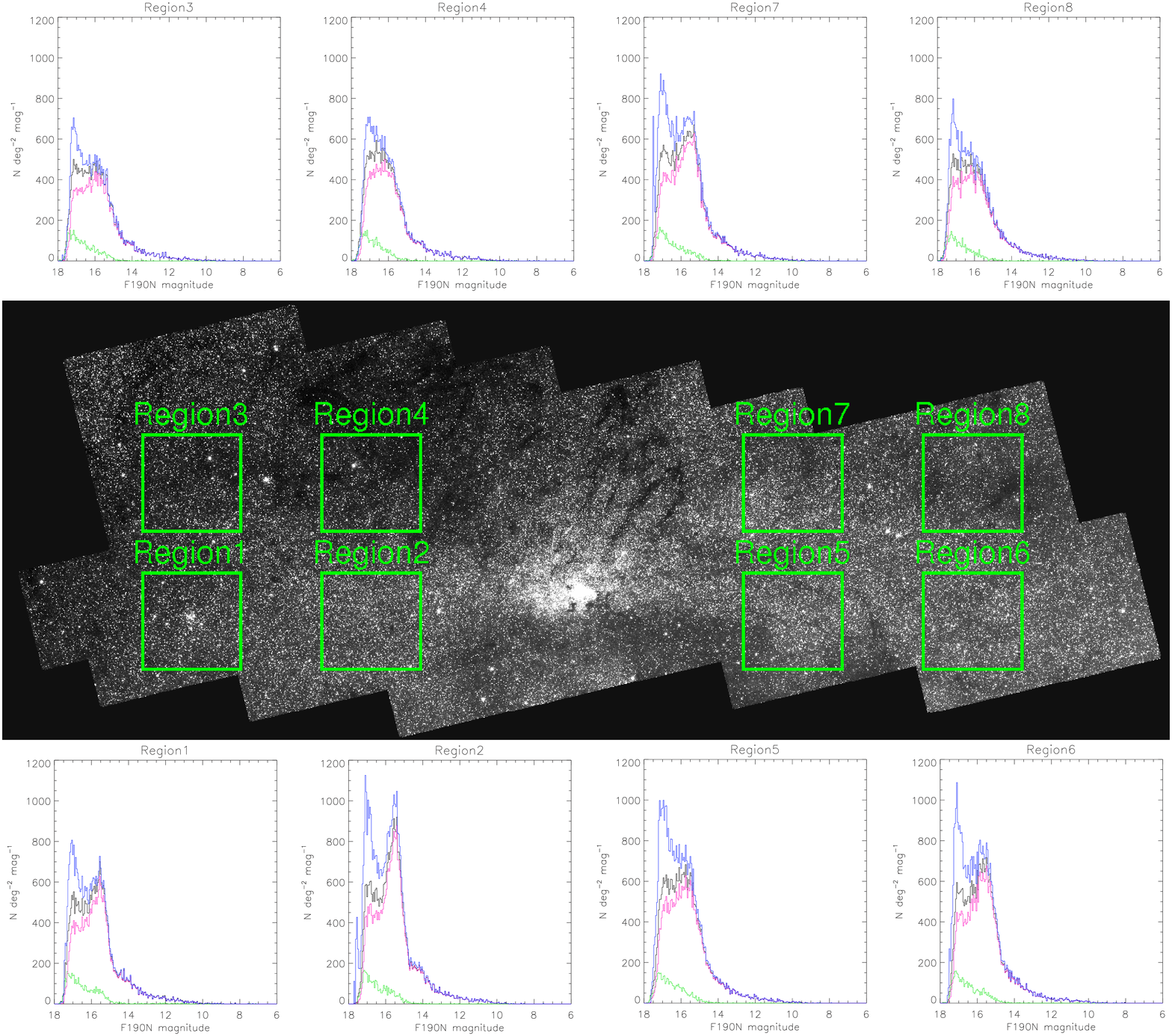,width=1.0\textwidth,angle=0}
    }
 \caption{Outlines of the regions in the F190N mosaic that are used
to extract the magnitude distributions shown in the individual
panels. The black line in each panel represents the original
magnitude distribution, while the green and pink lines are for the
foreground star and remaining (GC) contributions. The blue line is
for the GC distribution after the detection
          incompleteness correction.
 }
 \label{f:lum_dis}
\end{figure*}

Fig.~\ref{f:lum_dis} presents the magnitude distributions
constructed for eight different regions. The magnitude
distributions of the foreground stars in Regions 3, 4 and 8 show
shifts to the dim side because of the presence of foreground
molecular clouds. Similarly, the low star number density of Region
5 can be accounted for by the presence of the 50 km/s molecular
cloud (M-0.13-0.08). In the remaining 4 regions, the peak around
15.5 magnitude appears to become increasingly more prominent in
the regions closer to Sgr A* (cf. Region 1 and 2) and to the
Galactic Plane (cf. Region 6 and 7).

\subsection{Pa$\alpha$ emitting sources}\label{s:palpha}
In Fig.~\ref{f:total_color}, we mark the positions of the sources
that show significant Pa$\alpha$ emission. These sources are
defined from their large ratios of fluxes in the F187N and F190N
bands, corrected for local extinction ([12]). 79 of the 157
Pa$\alpha$ sources are apparently located in the three clusters,
including most of the evolved massive stars identified
spectroscopically (e.g., O If, LBV and WR stars). The remaining
sources appear to be distributed outside of these three clusters
or in the field. Two third of these field Pa$\alpha$ sources fall
in regions between the Arches/Quintuplet and Sgr A*, while the
other one third are located on the negative Galactic longitude
side. A small fraction of the field Pa$\alpha$ sources are clearly
associated with extended HII regions.

\subsection{Diffuse Pa$\alpha$ emission}\label{s:diffuse}
\begin{figure*}[!thb]
  \centerline{
    \epsfig{figure=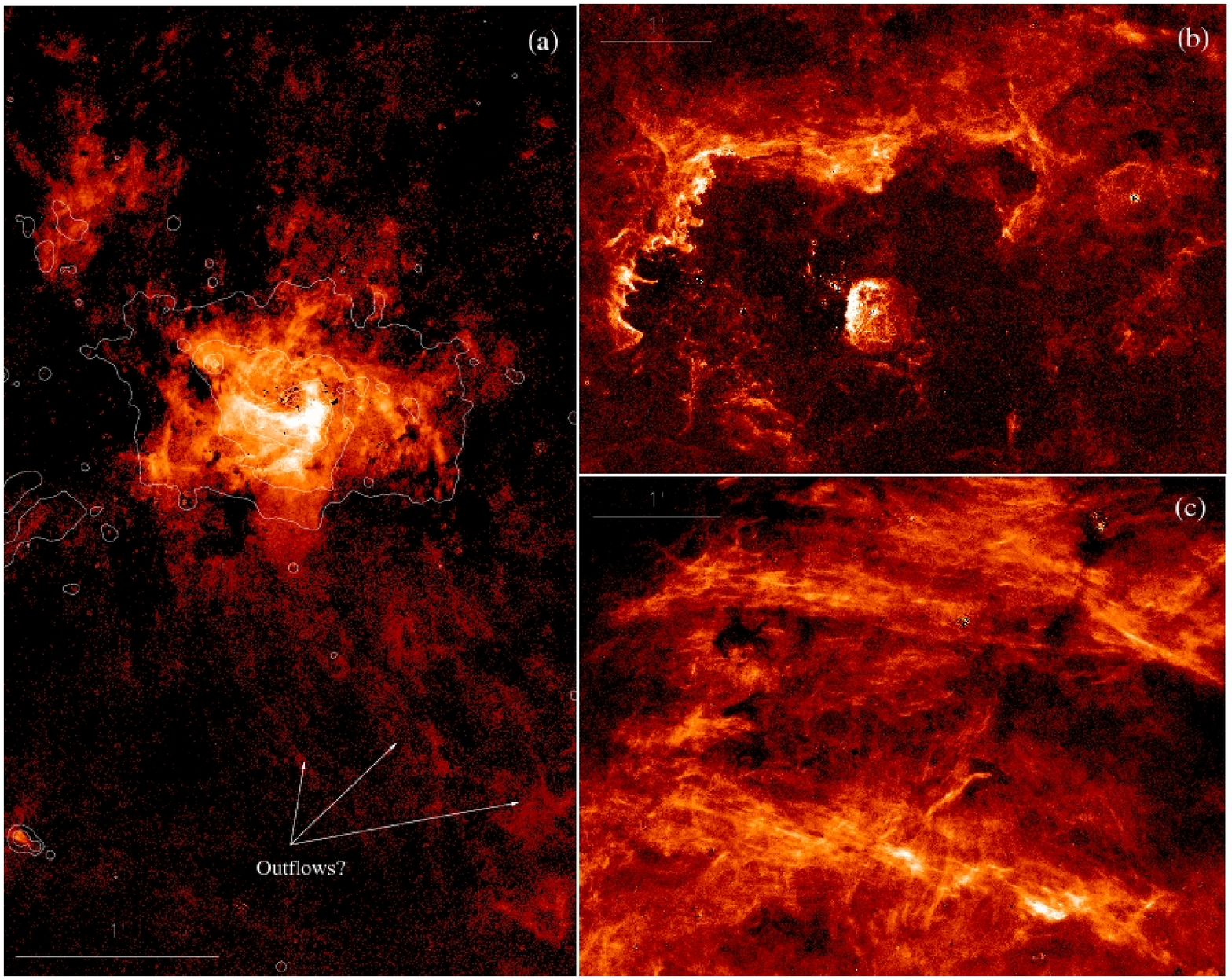,width=0.6\textwidth,angle=-90}
    }
 \caption{Close-ups of distinct diffuse P$\alpha$ features: (a) the central
region around Sgr A*, with the overlaid \irac\ 8\micron\ intensity
contours at 1,3,10, and 30  $\times 10^3 {\rm~mJy~sr^{-2}}$; (b)
the Sickle nebula; (c) Thermal Arched filaments. All are projected
in the Galactic coordinates.
 }
 \label{f:three}
\end{figure*}

Fig.~\ref{f:three} presents three sample regions that show
distinct diffuse Pa$\alpha$ emission features at the full
resolution. Around Sgr A*, one can clearly see a bright swirling
structure, as was discovered by Scoville et al. 2003 ([14]). But
in addition, Fig.~\ref{f:three}a shows low surface brightness
filaments which are nearly perpendicular to the Galactic Plane,
which may be an indication for outflows from the very central
region around Sgr A*. The well-known radio-thermal filaments in
the Sickle nebula and Thermal Arc are now resolved into many thin
filaments, which are apparently illuminated by the Arches and
Quintuplet clusters (Fig.~\ref{f:three}b,c). The Sickle nebula, in
particular, shows several finger-like structures similar to those
seen in M16. But there are also many fuzzy, low surface brightness
structures between the Quintuplet cluster and the bright rims. On
the Galactic west side of the Sickle nebula, one can also see a
new ring-like Pa$\alpha$ nebula. Follow-up spectroscopic
observations (Mauerhan et al 2010, in preparation) show that the
central Pa$\alpha$ source is an LBV star, similar to the Pistol
star.

\begin{figure*} %[t]
 \centerline{\epsfig{figure=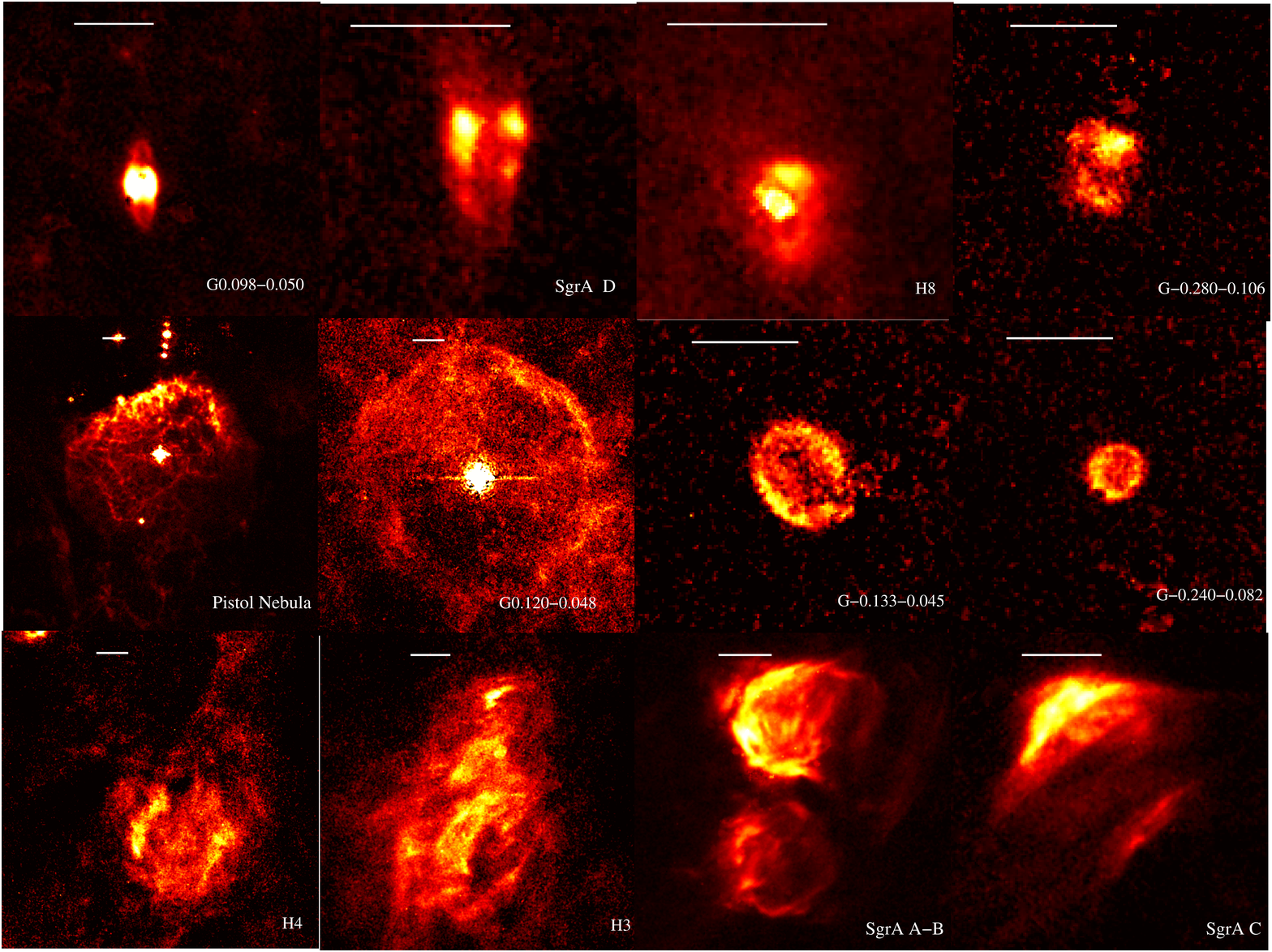,width=0.7\textwidth,angle=0}}
\caption{%\footnotesize
Close-ups of various compact Pa$\alpha$ nebulae
(equatorially projected). Detected sources have been subtracted,
except for identified bright Pa$\alpha$ emission stars that
appear to be central sources of the nebulae. Approximate Galactic
coordinates are used as labels of the nebulae, except those with well-known
names. The bar in each panel marks a 0.2 pc scale at the GC distance.}
\label{f:fig5}
\end{figure*}

Our survey reveals many new Pa$\alpha$ nebulae too. For example,
on the negative Galactic longitude side, we find two new extended
HII regions at (l,b)=(-0.13,0.0) and (-0.28,-0.036). Compared to
the Thermal Arc, they are much dimmer. That is why these regions
do not stand out against the strong background in the \spitzer\
\irac\ 8 $\mu$m PAH image. Another interesting discovery is the
existence of various linear filaments in the two H II regions.
These filaments probably trace local magnetic field. We give the
full resolution close-ups of extended Pa$\alpha$ emission in
Fig.~\ref{f:fig5}. The upper row shows the interaction between the
stars and its surrounding ISM. Nebulae in the middle row are
probably due to stellar ejecta. Nebulae in the bottom row are
several structures that are known as thermal radio sources and are
now well resolved. For example, Sgr A-A and Sgr A-C exhibit clear
bow shock structures.

\section{Discussion}\label{s:discussion}
\subsection{How do the massive stars shape the ISM?}\label{s:shape}
Intense ionizing radiation from the three massive star clusters
clearly illuminates, erodes and destroys nearby molecular clouds.
Our Pa$\alpha$ mosaic provides examples for all these three
processes. As shown in Section~\ref{s:diffuse}, the thin filaments
are brighter toward the Arches and Quintuplet clusters, which
presents the direct evidence that they are ionizing the surface of
the surrounding giant molecular clouds. The finger-like structures
at the surface of the Sickle nebula hint that the radiation from
the Quintuplet cluster is shaping the ISM, while the fuzzy
Pa$\alpha$ features between the Sickle nebula and Quintuplet
clusters probably represent remnants of molecular clouds that have
already been largely dispersed.

\subsection{What is the origin of the field Pa$\alpha$
sources}\label{s:origin} We find that many  of the
Pa$\alpha$-emitting stars are located outside of the three young
massive star clusters. As we mentioned previously, they are
evolved massive stars with ages of a few Myr. Studying their
origin can help us understand the star formation mode within the
\gc.

Two third of these field Pa$\alpha$ stars are located on the
positive Galactic longitude side of Sgr A* and are close to the
three massive star clusters. Stolte et al 2008 ([7]) have found
that the Arches cluster is moving toward the Galactic east in a
direction parallel to the Galactic plane. In
Fig.~\ref{f:total_color}, one can see that there are several
Pa$\alpha$ sources that appear in the opposite direction of the
motion. So one possible scenario is that these Pa$\alpha$ sources
are ejected from the clusters. Because of the mass segregation,
massive stars tend to sink into the cluster centers, where
three-body encounters are frequent, leading to ejections of stars
and formation of tightly bound binaries. So some of the Pa$\alpha$
sources could be due to the dynamic ejection, although detailed
simulations are needed to determine the efficiency of the process
and the distribution of such ejected stars. X-ray observations
have further shown the presence of three bright X-ray point
sources in the Arches cluster ([15] and reference therein). These
sources have hard thermal spectra, which are most likely due to
colliding winds in individual binaries [15]. Assuming that these
three X-ray sources represent all tightly bound binaries of
massive stars with strong winds, one may conclude that there
should be only a few massive stars at most that may be kicked out
from the cluster. Stolte et al 2008 ([7]) also suggested that at
most 15\% of the total mass may have been stripped away from the
Arches cluster by the tidal force in the GC. Due to the mass
segregation, the fraction of the massive stars stripped from the
cluster should be even smaller. Therefore, the cluster may not be
a primary source of the field Pa$\alpha$ sources. Most of the
field Pa$\alpha$ sources probably formed in isolation or in small
groups, independent of the Arches or Quintuplet cluster.

\subsection{Current star formation process}\label{s:current}

Fig.~\ref{f:total_color} shows that some of the field Pa$\alpha$
sources are associated with extended Pa$\alpha$ diffuse nebulae, which
represent regions of massive star formation in small groups.
These regions cannot be too young because of the presence of the evolved
massive stars represented by the Pa$\alpha$ sources.
The star groups also cannot be very rich, because of substantially
fewer such sources
than in the Arches cluster, which has more than 10 $Pa\alpha$-emitting sources
within 10" radius. Another evidence for the low mass of the groups is the
small sizes and relatively low diffuse Pa$\alpha$ intensities of the nebulae.

Our survey also for the first time unambiguously reveals the
presence of two large-scale HII complexes  on the negative
Galactic longitude side, These complexes, together with the
presence of tens of Pa$\alpha$ sources strongly indicates recent
star formation on this side of the Galactic disk. Overall,
however, the star formation is substantially less active than that
on the positive side. One possibility is that the star formation
on the negative side preceded the positive one; i.e.,
clusters/groups have largely been resolved.

Additional evidence for the star formation on the negative side
comes from radio observations. Law 2010 ([16]) have shown that a
giant Galactic Center lobe of size $\sim$100 pc is not centered at
the \gc, but shifts to the negative side. This shift may be due to
a strong starburst on this side about several Myr ago, responsible
for the lobe. Clearly, more observations and detailed modelling
are required to further the understanding of the global recent
star formation pattern in the GC.

\subsection{Star formation history}\label{s:intermediate}

\begin{figure*} %[t]
 \centerline{\epsfig{figure=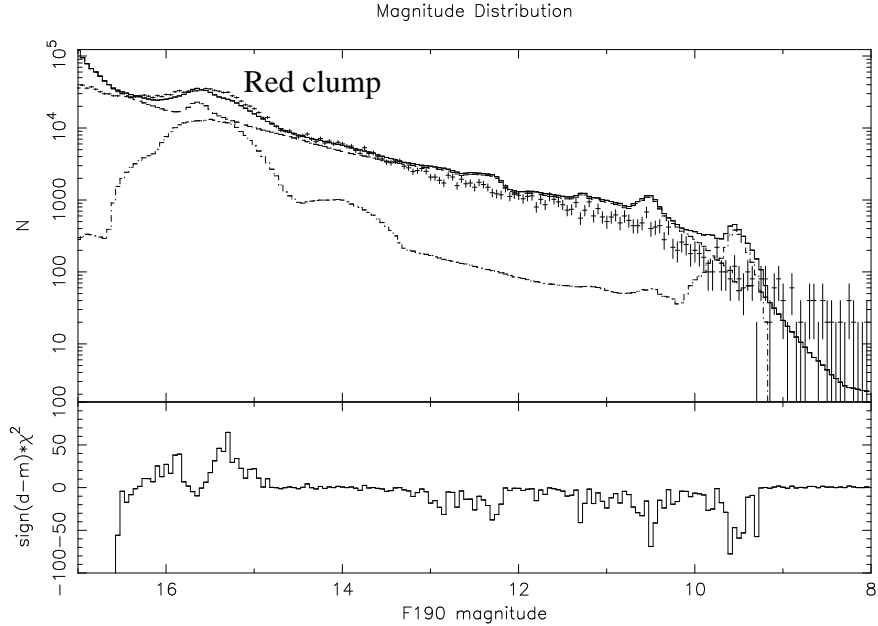,width=0.8\textwidth,angle=0}}
\caption{%\footnotesize
Top: the fitting result.~The three lines from bottom to up are the
recent star formation activity which began $\sim$350 Myr ago, the
continuous star formation through the whole history of the
Universe and their combination. Bottom: the fitting residuals }
\label{f:model}
\end{figure*}

Fig.~\ref{f:model} compares the magnitude distribution in
1.9$\mu$m with a stellar population synthesis model. This
comparison shows that the overall shape of the distribution can be
reasonably modelled with a constant star formation rate, plus a
major burst about 350 Myr ago. The large deviation of the
distribution from the model at the small magnitude end is probably
caused by a problematic treatment of TP-AGB phase, which has been
calibrated with stellar clusters in the Magellanic Clouds. This
evolutionary phase may be sensitive to the metallicity, which
could be substantially different in the GC from the calibration
clusters. The burst in the magnitude distribution model is
required to explain the presence of the m$_{1.90 \mu m}\sim15.5$
peak in the data, representing red clump stars with initial mass
around 2 $M_{\odot}$. The magnitude position of this peak varies
among different lines-of-sight (Fig.~\ref{f:lum_dis}), which hints
that this component is indeed in the \gc , not the spiral arms.
This spatial variation of the peak intensity further suggests that
the distribution of the red clump stars may follow a disk-like
structure. While a detailed investigation of the magnitude
distribution is yet to be made, it is clear that it can be used to
shed important insights into the star formation history in the GC.

\section{Summary}\label{s:summary}
Based on our \hst\ Pa$\alpha$ survey of the \gc, we have detected
0.6 million stars and mapped out many extended Pa$\alpha$ emission
features. The main results from our preliminary analysis of these
products are summarized in the following:

\begin{itemize}
\item We have identified 157 Pa$\alpha$-emitting sources.
They are most likely evolved massive stars and trace very recent
star formation in the \gc. About half of the sources are located
outside the three well-known massive star clusters. These field
sources formed probably mostly in small groups. Some of the sources
are still associated with relatively compact H~II regions.

\item The diffuse Pa$\alpha$ map allows us to resolve many fine structures
in known large-scale thermal radio features in the \gc.
These structures represent various stages of
the interplay between massive stars and their ISM environment,
including stellar mass ejection as well as the ionization and destruction
of nearby molecular clouds. The filamentary morphology of some of the
structures further indicates that magnetic
field plays an important role in shaping the ISM in the GC.

\item We have clearly detected two HII complexes as well as
$\sim$20 Pa$\alpha$ sources on the negative Galactic longitude
side, suggesting recent star formation there, though probably
earlier and/or weaker than that on the other side. The star
formation may be linked to be partly responsible for the GC lobe,
as identified by Law et al 2010 ([16], and reference therein).

\item We have found evidence for a starburst about 350 Myr years
ago, as characterized by an enhanced number of red clump stars
evolved from initial masses $\sim$2 $M_{\odot}$. These stars seem
to be primarily located in a disk-like region around Sgr A*.
\end{itemize}

These preliminary results demonstrate that near-IR surveys can be
used to significantly advance our understanding of the mode and
history of star formation in the GC, shedding insights into what
may occur in nuclear regions of other galaxies.

\end{document}